# Rise of post-pandemic resilience across the distrust ecosystem

Lucia Illari[1,2], Nicholas J. Restrepo[3], and Neil F. Johnson[1,2,*]
[1]Dynamic Online Networks Laboratory, George Washington University, Washington, D.C., 20052, U.S.A.
[2]Physics Department, George Washington University, Washington, D.C., 20052, U.S.A.
[3]ClustrX LLC, Washington, D.C., U.S.A.             * corresponding author

**Abstract:**

**Why is distrust (e.g. of medical expertise) now flourishing online despite the surge in mitigation schemes being implemented? We analyze the changing discourse in the Facebook ecosystem of approximately 100 million users who pre-pandemic were focused on (dis)trust of vaccines. We find that post-pandemic, their discourse strongly entangles multiple non-vaccine topics and geographic scales both within and across communities. This gives the current distrust ecosystem a unique system-level resistance to mitigations that target a specific topic and geographic scale -- which is the case of many current schemes due to their funding focus, e.g. local health not national elections. Backed up by detailed numerical simulations, our results reveal the following counterintuitive solutions for implementing more effective mitigation schemes at scale: shift to 'glocal' messaging by (1) blending particular sets of distinct topics (e.g. combine messaging about specific diseases with climate change) and (2) blending geographic scales.**

## Introduction

Distrust and its associated mis/disinformation -- however defined -- is now a widespread threat to public health (e.g., abortion, COVID-19, monkeypox), science (e.g., climate change), election processes and even national security (*1–5*). The pandemic made this worse with many people turning to their trusted online communities during social distancing in order to get advice, and to share any distrust of official health messaging (*6–18*). Within a month of the U.S. national emergency declaration (*19*), Facebook -- the largest and most widely used social media platform -- saw a 50% increase in messaging and 70% increase in time spent (*20*), driving its monthly active users to 2.6 billion.

To combat this, myriad ingenious mitigation strategies have been introduced and now implemented online (*12, 21–38*). Depending on a mitigation scheme's funding source, its focus at any time is typically a topic such as COVID-19, or abortion, or monkeypox, or climate change, or elections, while its geographical focus is at the level of a specific state, or the national scale (e.g., APS), or worldwide (e.g., E.U.) But given so many diverse mitigation schemes, why is distrust still so widespread?

Here we provide an answer to this question, and the counterintuitive solution that this answer suggests. Specifically, we show that post-pandemic distrust has developed a massive glocal web that -- within individual communities and across interconnected communities -- blends distinct topics, locations and geographic scales. This makes it resilient to current mitigation schemes that only focus on a specific topic or geographic scale (e.g. due to funding mandate) (*39*). Given that such schemes also operate independently, this suggests widespread distrust would remain resilient even if these schemes were implemented at mass scale. We show this in Figs. 1-3 by analyzing the post-pandemic discourse across the Facebook ecosystem of approximately 100 million individuals that -- pre-pandemic -- was centered on vaccine distrust (*40*). Combining this with an agent-based simulation, Fig. 4 shows how this web-of-distrust can be dismantled by making individual mitigation schemes blend topics and scales.

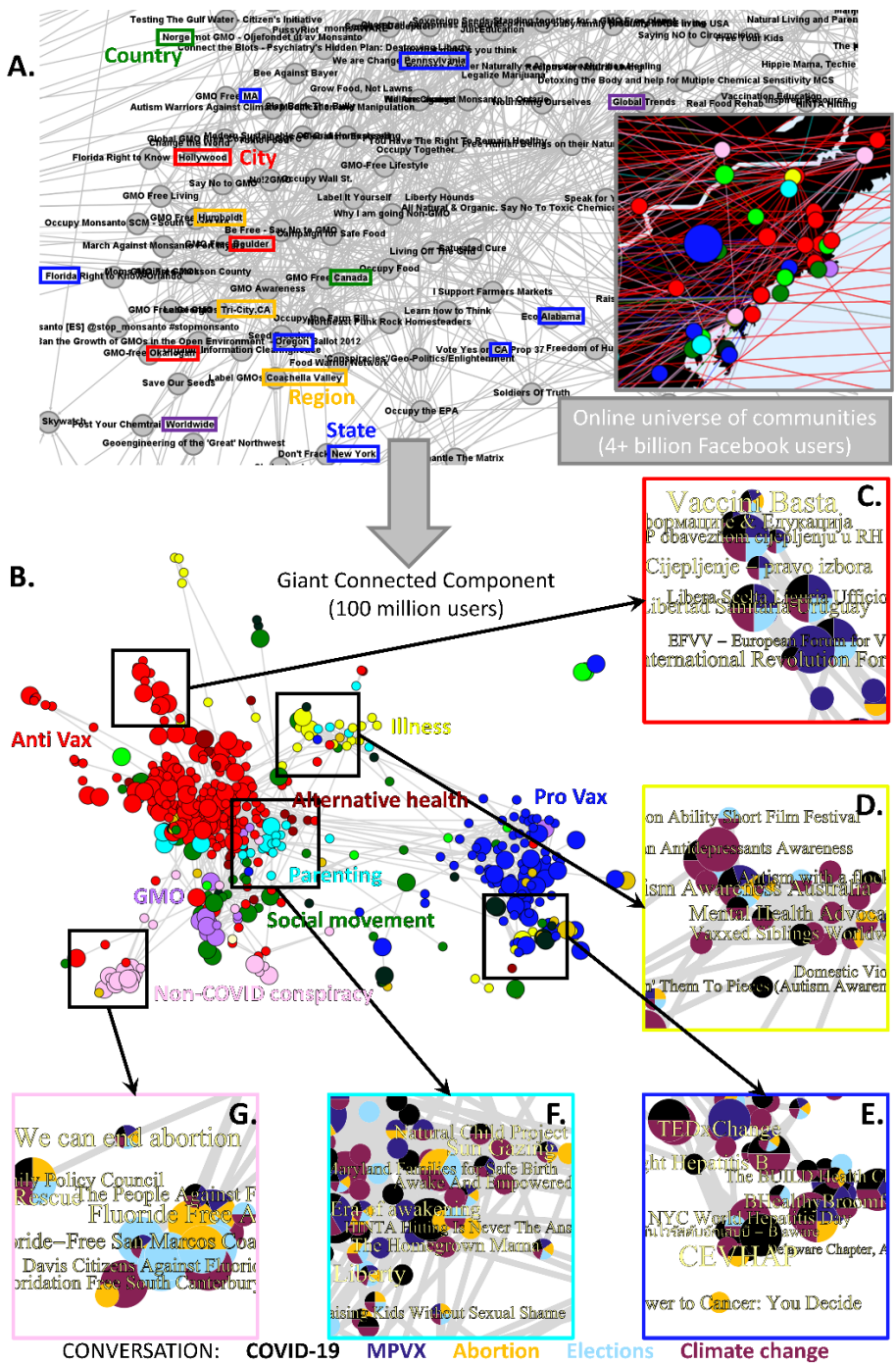

**Figure 1: Post-pandemic distrust entangles topics, locations, and geographical scales. A:** Illustrative sample of our data *(40)*. Each circle is a Facebook community (page). Communities promote page-level links to each other. Inset illustrates communities' locations in U.S. north-east. **B:** Giant connected component of communities classified according to their stance on vaccines. Neutral communities (i.e. non-blue, non-red) are subclassified by their primary interest, e.g. parenting (light blue). Node size indicates geographic scale: large nodes are local communities, small nodes are global ones. **C-G:** Each community's discourse sub-classified by proportion of dominant topics (black is Covid-19, dark purple is monkeypox (MPX), gold is abortion, light blue is elections, dark magenta is climate change). SI Sec. 6 gives full information and shows complete network.

**Methods**

To examine how the distrust discourse changed post-pandemic, we revisited the 2019 Facebook ecosystem from Ref. (*40*) that had centered around vaccines and comprised interlinked anti-, pro-, and neutral-vaccination Facebook pages. Our full methodology is given in the SI and follows Ref. (*40*). Each Facebook page is a community (i.e. node in Fig. 1) with a unique ID, and has nothing to do with community detection in networks. These communities provide spaces where users gather around shared interests, thereby promoting trust among them (*41–46*) and potential collective distrust of other issues (*6*, *22*, *40*, *47–54*). Our trained researchers manually and independently classify each community involved in the vaccine debate, with subsequent consensus checks performed in cases of disagreement. This yielded a network of 1356 interlinked communities across countries and languages, with 86.7 million individuals in the largest network component; 211 pro-vaccination communities (blue nodes, Fig. 1B) with 13.0 million individuals; 501 anti-vaccination communities (red nodes, Fig. 1B) with 7.5 million individuals; 644 neutral communities (non-blue or red nodes, Fig. 1B) with 66.2 million individuals. These neutral communities were further sub-categorized by type based on their title and description (e.g. parenting). The discourse within each of the 1356 communities was categorized by topic prevalence. Though topics outside the five dominant ones (COVID-19, MPX, abortion, climate change, and elections) are mentioned, their frequency is generally much lower (e.g. sports).

A link is shown between two communities (Facebook pages) *A* and *B* when community *A* recommends community *B* to its members at the page level. This creates a prominent hyperlink from *A* to *B* indicating community *A*'s interest in *B*, which is different than if a member of *A* had simply mentioned some content from *B*. A link does not necessarily mean the two communities agree. Instead, it directs the attention of *A*'s members to *B*, and vice versa it exposes *A* to feedback and content from *B*. While not all members will necessarily pay attention, a committed minority of 25% can be enough to influence the stance of an online community (*55*).

Of the 1356 communities, 342 identify as local and have around 3.1 million individuals, while the remaining 1014 communities are global with around 83.7 million individuals. A global community is a page with broad, worldwide focus that is not tied to a specific location, while a local community is focused on a specific geographic area, such as a neighborhood, city, county, state, or country. For example, a community with the name "Vaccine information for Los Angeles County parents" is local since it mentions a specific location, while a page with the name "Vaccine information for parents" or "Global Trends" is global as it implies worldwide focus. Similarly, a page can be classified as global in the topic space if it has a broad focus, or local if it refers to a narrow focus, e.g. pages that discuss specifically and only elections, or abortion-and-elections. The size of each community can be estimated by the number of likes, given that the average user only likes one Facebook page on average (*7*)—however, our analysis and findings are not dependent on this.

**Results**

Figure 1 shows this web-of-distrust during the period 5/1/2022 to 10/17/2022, which included several significant events: (1) the first confirmed U.S. case of the monkeypox outbreak (*56*); (2) the U.S. Supreme Court's reversal of Roe v. Wade (*57*); (3) President Biden signing into law the Inflation Reduction Act (*58*); (4) primary and run-off elections ahead of the November midterms (*59*). Communities that share more links appear visually closer together and have a higher likelihood of exerting influence on each other through shared content and infiltration. This is because the layout results from a color-agnostic physical calculation (ForceAtlas2) in which nodes repel each other with a force that decays with separation, and linked nodes have an additional attractive spring force (*60*).

This web-of-distrust entangles 5 dominant topics within and across communities: abortion, monkeypox (MPX), COVID-19, climate change, elections (panels C-G). This means that within a given community, distrust in one topic can immediately be reinforced by distrust of another topic(s) as well as by the collective distrust of other communities to which it is linked. Hence post-pandemic intervention on one topic and scale invites pushback from distrust on other topics and across scales, making the distrust ecosystem resilient to mitigation schemes that focus on a particular topic and geographic scale. For example, distrust of state elections within a community associated with a small U.S. city is reinforced by distrust in the U.S. federal government on monkeypox, which is then reinforced by distrust over climate change from another community representing itself as the U.K. mainstream. This global-local entanglement across topics and scales adds new resilience to the distrust ecosystem.

Figure 1 has counterintuitive implications for messaging the public and hence intervening against distrust (see SI for details and statistical analyses). One might expect discussions surrounding elections and abortion to focus on a specific geographical scale within the U.S. due to specific laws, politics, and healthcare systems. However, a chi-square test shows the opposite is true. Among communities only discussing a single topic, 25.21%, 35.14%, 17.65%, 38.71%, and 24.32% that discuss COVID-19, MPX, abortion, elections, and climate change, respectively, are local—in contrast to the expected 31.9% if there were no correlation. This implies that any single-topic messaging around COVID-19, or abortion, or climate change, should have more of a global perspective, while that around MPX and elections should be more local. This difference between MPX and COVID-19 also warns against one-size-fits-all for public health messaging, despite both being emerging diseases.

Moreover, traditional anti-vax communities—a third of which are local—have a lower interest in abortion (Fig. 1C). Climate change is the most popular topic among illness communities, of which only a few are local (Fig. 1D). Parenting communities are closely associated with alternative health communities and discuss many topics (Fig. 1F). Conspiracy theory communities show high interest in abortion and elections and are 50% local (Fig. 1G). GMO communities have the highest percentage of local members (53.8%) as compared to all communities in Fig. 1B for which only 25.2% are local.

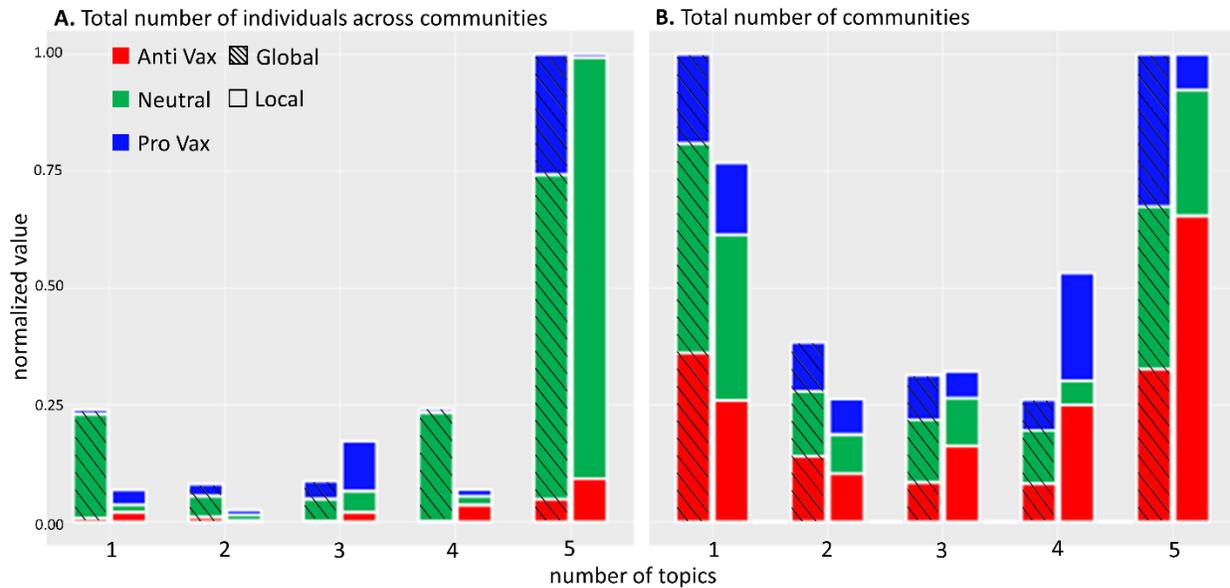

**Figure 2: The numbers of topics discussed by (A) individuals, and (B) communities. Categorized by their geographic scale and stance in the vaccine debate. Global communities are shaded. Results normalized by the maximum value. SI Sec. 9 confirms that these patterns do not arise by chance.**

Figure 2 confirms the wide distribution across topics and geographic scales of the distrust discourse. For anti-vaccination communities, this is the number of topics about which distrust is being actively promoted. One might expect that as a community addresses more topics, the number of potential flashpoints for internal disagreements would increase—hence there would be less communities and individuals engaging in higher numbers of topics—but the opposite happens. Furthermore, local communities are overrepresented in topics compared to the full dataset: hence future mitigation schemes, including global ones, should be designed such that local communities and their interests feature in a prominent way.

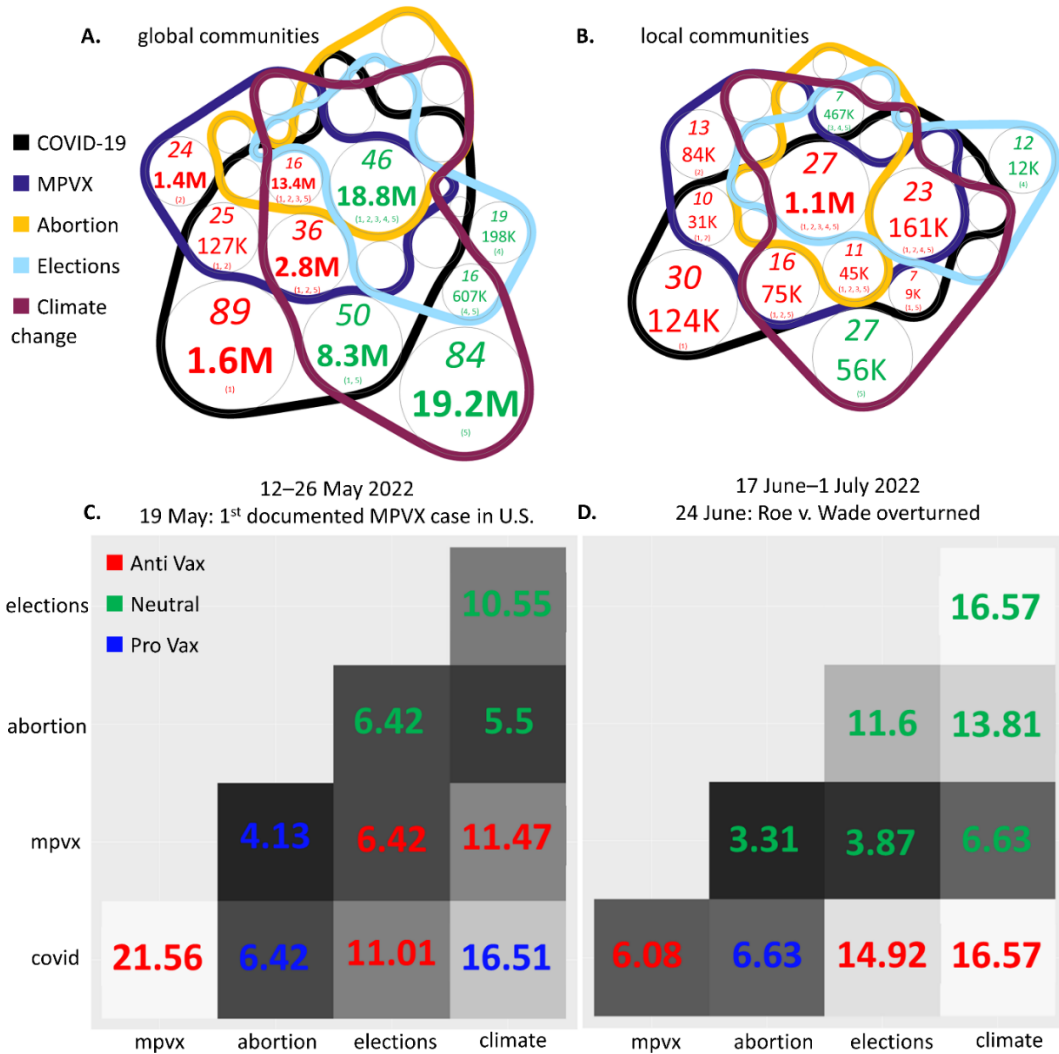

**Figure 3: Topic mix within (A) global communities, and (B) local communities, shown using n-Venn diagrams (see SI Sec. 8 for simple example explaining an n-Venn diagram (75)). Number of communities in italics at top. Number of individuals in middle (in bold if >1M). Specific combination of topics at bottom. Number is shown in red (or green) if there is a prevalence of anti-vaccination (or neutral) communities. Only regions with >3% of total communities are labeled. C-D: Number of communities discussing two topics for (C) 12 May 2022 to 26 May 2022 around first documented case of monkeypox in U.S. (19 May); (D) 17 June 2022 to 01 July 2022 around 24 June overturning of Roe v. Wade. Black-to-white scale represents low-to-high counts of both the topics in discourse. Text color indicates vaccine stance of the majority of communities that discuss the two topics. Symmetric squares are left empty.**

Figures 3A and 3B show the importance of specific combinations of topics for effective public messaging. The subset of communities discussing all five topics is the third largest but has the highest number of individuals at 19.9 million. There are fewer communities that only discuss MPX, abortion, or elections. There is significant conversation overlap between COVID-19 and climate change, particularly in global communities. Despite the involvement of pro-vaccination communities in these discussions, the dialog is mostly led by communities that do not promote guidance consistent with current scientific consensus—and in many cases, these are communities that actively oppose it, especially at the local level.

Figures 3C and 3D compare how different pairs of topics featured across the distrust ecosystem during key 2022 periods. In Fig. 3C which includes the first U.S. MPX case, anti-vaccination communities quickly took control of the MPX conversation by connecting it with COVID-19. The pro-vaccination communities remained stuck on climate and COVID-19. This suggests that there was a lack of professional medical guidance about monkeypox during this critical period. In Fig. 3D which includes the reversal of U.S. abortion law Roe v. Wade, the pro-communities were able to dominate the abortion and COVID-19 discussion whereas neutrals had the upper hand in the abortion and climate conversation—but the antis stayed with "COVID-19 and climate" and "COVID-19 and elections." This suggests a phenomenon akin to real multi-virus interference (*61*): messaging combining a few dominant topics helps suppress distrust in other topics.

Figure 4 uses an agent-based simulation to compare the effectiveness of mitigation schemes that target a specific topic and geographic scale (like most current schemes) to ones that combine topics and geographic scales. The model allows communities to reactivate after mitigation: in Fig. 4A, reactivation probability is determined by the ratio of global communities followed to all communities followed; in Fig. 4B, reactivation probability for a specific topic is determined by the ratio of communities followed that discuss that topic to all communities followed. Communities that follow only local communities or do not follow any communities discussing the topic have a 0% chance of reactivation, while those following only global communities or all communities discussing the topic have a 100% chance of reactivation.

Figure 4A shows that eliminating distrust in all local nodes is not possible—even with official messaging that specifically targets the local level, i.e. it is not possible to 'silence' all local communities (nodes) all the time or all at once. It also demonstrates how the network's high interconnectedness makes it resistant to messaging that only targets one level. Figure 4B demonstrates explicitly the superiority of multi-topic vs. single-topic messaging for de-bunking and pre-bunking. The curves in Fig. 4B show how the number of communities discussing a particular topic decrease over time due to the mitigation strategy, and the overall trends are represented by averaged reduction curves. Separate curve components that were averaged can be found in SI Sect. 11. The effectiveness of the type of messaging was measured by the rate and magnitude of the decrease in the number of communities over time. There is an overall decrease despite the implementation of a reactivation strategy. Even for two-topic messaging, the proportion of communities (nodes) discussing the topic is reduced to less than 50% in less than 600 steps, and this reduction is even faster with more topics. Though the results for 3, 4, and 5-topic messaging are similar, these results show that just 2-topic messaging is already highly effective and hence should be employed de facto in all future mitigation schemes—e.g. COVID-19 and climate change (see heat maps). Another practical advantage is that complete knowledge of the web-of-distrust is not required in order to implement this multi-topic approach effectively.

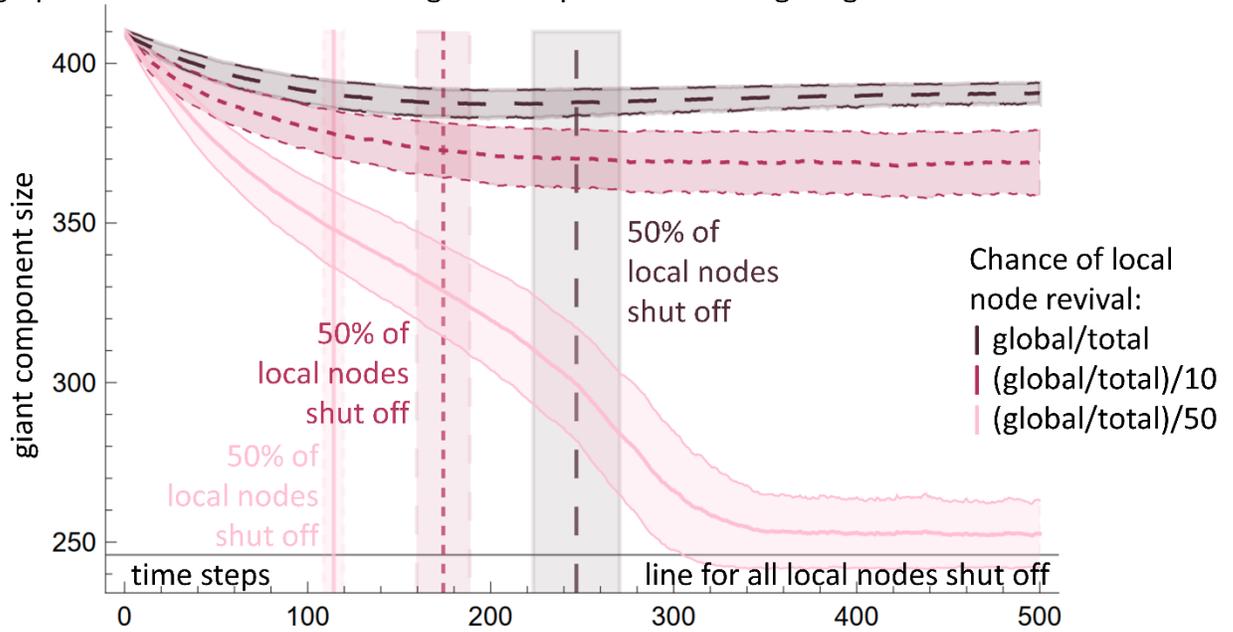

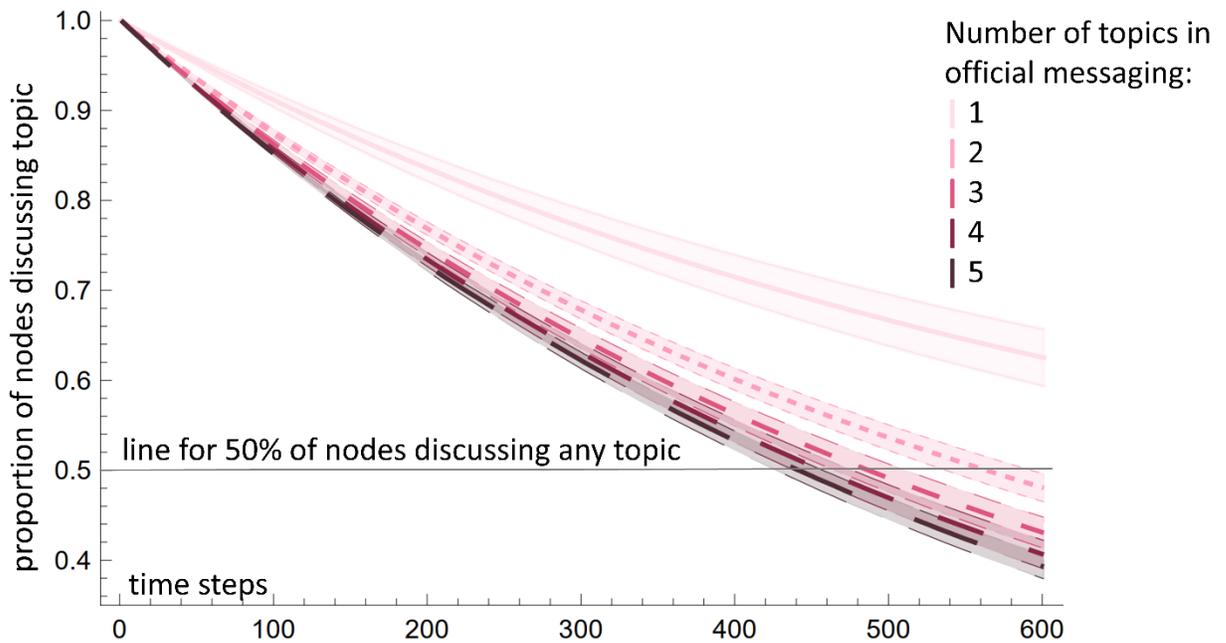

**Figure 4: Agent-based simulation results for mitigation schemes. (A)** Global-local cohesion of the system: giant component size vs. time. Vertical lines indicate $\mu \pm \sigma$ when 50% of local nodes are silenced. Top two curves level off, while bottom curve dips before leveling off just above line indicating giant component size when all local nodes are silenced. **(B)** Single vs multi-topic debunking: proportion of nodes discussing $n$ topics versus time. Results show much higher impact for multi-topic messaging.

## Discussion

Our study obviously has limitations. Our analysis does not yet stretch to all languages (*62*): but SI Sect. 2 uses page administrator locations as a proxy to show that our dataset is indeed diverse. Additional sentiment analysis or natural language processing of all posts could also be conducted, and of course other social media exist. We note that although our study is technically a large sample of the actual online population, the large number involved (approximately 100 million) suggests it qualifies as a crude population-level map. Indeed, we did not obtain the nodes and links by simple sampling but rather by detection and then following links from node to node. After a while, this tended to return to the same nodes and hence, like circling the globe, hints that we have mapped—albeit crudely—the skeleton of the true online distrust ecosystem.

Overall, therefore, we believe that we have provided a crude mapping of the online distrust ecosystem on Facebook. Since Facebook is the largest social media platform, it is plausible that the cohort of approximately 100 million users is therefore a crude representation of the entire population. Specifically, we analyzed the changing discourse in the Facebook ecosystem of approximately 100 million users who pre-pandemic were focused on (dis)trust of vaccines. We found that post-pandemic, their discourse strongly entangles multiple non-vaccine topics and geographic scales both within and across communities -- hence giving the current distrust ecosystem a unique system-level resistance to mitigations that target a specific topic and geographic scale. Since this is the case of many current schemes due to their funding focus, e.g. local health not national elections, this study raises questions about the current approach to funding of distrust mitigation research. Our results provide the following counterintuitive solutions for implementing more effective mitigation schemes at scale: shift to 'glocal' messaging by (1) blending particular sets of distinct topics (e.g. combine messaging about specific diseases with climate change) and (2) blending geographic scales.

**Data availability and code availability**
All data needed to evaluate the conclusions in the paper are present in the paper and Supplementary Information (Refs. (*40, 42, 55–59, 63–70*)). The code used to generate the map in Fig. 1, and from which the results in Figs. 2-3 are obtained, is Gephi which is free open-source software. Figure 4 was obtained using Mathematica.

**Author contributions:** L.I. analyzed the results and generated the figures. N.F.J. supervised the project. L.I and N.F.J. wrote the paper. All authors were involved in reviewing the final manuscript, and in the conceptualization, methodology, and validation.

**Competing interests:** The authors have no competing interests, either financial and/or non-financial, in relation to the work described in this paper.

**Acknowledgements**
N.F.J. is supported by U.S. Air Force Office of Scientific Research awards FA9550-20-1-0382 and FA9550-20-1-0383.

**All correspondence and material requests should be addressed to** N.F.J. neiljohnson@gwu.edu

**SUPPLEMENTARY INFORMATION (SI) CONTENTS:**

<u>Section 1</u>: Methodology:
Collecting data, building network, use of term "glocal", and analysis in the paper.

<u>Section 2</u>: Breakdown of page admin country locations. Color scheme for neutral nodes in network plots. The system before COVID-19, one year later, and three years later. Geolocalized network.

<u>Section 3</u>: Classification of neutral nodes.

<u>Section 4</u>: Example of Facebook banners promoting best-science Covid-19 guidance. Positions in network of the nodes that receive Facebook banners promoting best-science Covid-19 guidance.

<u>Section 5</u>: ForceAtlas2 layout and analysis showing dependence of layout on strength of bonding.

<u>Section 6</u>: Topic filter. System filtered by topic. System in October 2022 without node labels, and only with those labels appearing in Fig. 1.

<u>Section 7</u>: Chi-square test for topic-glocality.

<u>Section 8</u>: Use of nVenn diagrams. General system dynamics, glocal breakdown, and comparison to simulation.

<u>Section 9</u>: Relationship between the number of topics and glocality.

<u>Section 10</u>: Details of the geographic scale simulation simulation.

<u>Section 11</u>: Details of the topic simulation.

## References

1. Nobel Prize Summit 2023. *NobelPrize.org*, (available at https://www.nobelprize.org/events/nobel-prize-summit/2023/).

2. TRUST SCIENCE PLEDGE CALLS FOR PUBLIC TO ENGAGE IN SCIENTIFIC LITERACY. *News Direct*, (available at https://newsdirect.com/news/trust-science-pledge-calls-for-public-to-engage-in-scientific-literacy-737528151).

3. American Physical Society Takes On Scientific Misinformation, (available at http://aps.org/publications/apsnews/202203/misinformation.cfm).

4. AAAS 2022 Annual Meeting: How to Tackle Mis- and Dis-information | American Association for the Advancement of Science (AAAS), (available at https://www.aaas.org/news/aaas-2022-annual-meeting-how-tackle-mis-and-dis-information).

5. *Beyond disinformation – EU responses to the threat of foreign information manipulation* (2023; https://www.youtube.com/watch?v=YJf2pZGe36Q).

6. D. S. Ardia, E. Ringel, V. Ekstrand, A. Fox, Addressing the Decline of Local News, Rise of Platforms, and Spread of Mis- and Disinformation Online: A Summary of Current Research and Policy Proposals. *SSRN Journal* (2020), doi:10.2139/ssrn.3765576.

7. H. Inc, Digital Trends - Digital Marketing Trends 2022. *Digital Trends - Digital Marketing Trends 2022*, (available at https://www.hootsuite.com).

8. G. Lappas, A. Triantafillidou, A. Deligiaouri, A. Kleftodimos, Facebook Content Strategies and Citizens' Online Engagement: The Case of Greek Local Governments. *Rev Socionetwork Strat*. **12**, 1–20 (2018).

9. K.-K. Kleineberg, M. Boguñá, Competition between global and local online social networks. *Sci Rep*. **6**, 25116 (2016).

10. A. Rao, F. Morstatter, K. Lerman, Partisan asymmetries in exposure to misinformation. *Sci Rep*. **12**, 15671 (2022).

11. Weight-loss injections have taken over the internet. But what does this mean for people IRL? *MIT Technology Review*, (available at https://www.technologyreview.com/2023/03/20/1070037/weight-loss-injections-societal-impact-ozempic/).

12. Getting Ahead of Misinformation. *Democracy Journal* (2023), (available at https://democracyjournal.org/magazine/68/getting-ahead-of-misinformation/).

13. R. DiResta, The Digital Maginot Line. *ribbonfarm* (2018), (available at https://www.ribbonfarm.com/2018/11/28/the-digital-maginot-line/).

14. Misinformation, Crisis, and Public Health—Reviewing the Literature – MediaWell, (available at https://mediawell.ssrc.org/?post_type=ssrc_lit_review&p=58936).

15. H. J. Larson, Blocking information on COVID-19 can fuel the spread of misinformation. *Nature*. **580**, 306–306 (2020).

16. E. Douek, Content Moderation as Systems Thinking, (available at https://harvardlawreview.org/2022/12/content-moderation-as-systems-thinking/).

17. E. Chen, K. Lerman, E. Ferrara, Tracking Social Media Discourse About the COVID-19 Pandemic: Development of a Public Coronavirus Twitter Data Set. *JMIR Public Health Surveill*. **6**, e19273 (2020).

18. A. Semenov, A. V. Mantzaris, A. Nikolaev, A. Veremyev, J. Veijalainen, E. L. Pasiliao, V. Boginski, Exploring Social Media Network Landscape of Post-Soviet Space. *IEEE Access*. **7**, 411–426 (2019).

19. S. Ghaffary, People are using Facebook more than ever during the coronavirus pandemic — but its business is still taking a hit. *Vox* (2020), (available at https://www.vox.com/2020/4/29/21241601/facebook-coronavirus-pandemic-users-advertising-growth-making-losing-money-users-q1-2020-earnings).

20. Keeping Our Services Stable and Reliable During the COVID-19 Outbreak. *Meta* (2020), (available at https://about.fb.com/news/2020/03/keeping-our-apps-stable-during-covid-19/).

21. What's Being Done to Fight Disinformation Online, (available at https://www.rand.org/research/projects/truth-decay/fighting-disinformation.html).

22. Managing the COVID-19 infodemic: Promoting healthy behaviours and mitigating the harm from misinformation and disinformation, (available at https://www.who.int/news/item/23-09-2020-managing-the-covid-19-infodemic-promoting-healthy-behaviours-and-mitigating-the-harm-from-misinformation-and-disinformation).

23. Government to relaunch 'Don't Feed the Beast' campaign to tackle Covid-19 misinformation – Society of Editors, (available at https://www.societyofeditors.org/soe_news/government-to-relaunch-dont-feed-the-beast-campaign-to-tackle-covid-19-misinformation/).

24. Get the facts on coronavirus. *Full Fact*, (available at https://fullfact.org/health/coronavirus/).

25. COVID-19 Information Center | Meta. *COVID-19 Information Center*, (available at https://about.meta.com/covid-19-information-center).

26. VERIFIED: UN launches new global initiative to combat misinformation. *Africa Renewal* (2020), (available at https://www.un.org/africarenewal/news/coronavirus/covid-19-united-nations-launches-global-initiative-combat-misinformation).

27. Restoring Trust in Public Health (2023), , doi:10.26099/j7jk-j805.

28. New USD10 Million Project Launched To Combat the Growing Mis- and Disinformation Crisis in Public Health. *The Rockefeller Foundation*, (available at https://www.rockefellerfoundation.org/news/new-usd10-million-project-launched-to-combat-the-growing-mis-and-disinformation-crisis-in-public-health/).

29. Navigating Infodemics and Building Trust during Public Health Emergencies A Workshop | National Academies, (available at https://www.nationalacademies.org/event/04-10-2023/navigating-infodemics-and-building-trust-during-public-health-emergencies-a-workshop).

30. Understanding and Addressing Misinformation About Science A Public Workshop | National Academies, (available at https://www.nationalacademies.org/event/04-19-2023/understanding-and-addressing-misinformation-about-science-a-public-workshop).

31. S. Mirza, L. Begum, L. Niu, S. Pardo, A. Abouzied, P. Papotti, C. Pöpper, "Tactics, Threats & Targets: Modeling Disinformation and its Mitigation" in *Proceedings 2023 Network and Distributed System Security Symposium* (Internet Society, San Diego, CA, USA, 2023; https://www.ndss-symposium.org/wp-content/uploads/2023/02/ndss2023_s657_paper.pdf).

32. J. Roozenbeek, S. van der Linden, B. Goldberg, S. Rathje, S. Lewandowsky, Psychological inoculation improves resilience against misinformation on social media. *Science Advances*. **8**, eabo6254 (2022).

33. S. van der Linden, A. Leiserowitz, S. Rosenthal, E. Maibach, Inoculating the Public against Misinformation about Climate Change. *Global Challenges*. **1**, 1600008 (2017).

34. Under the surface: Covid-19 vaccine narratives, misinformation and data deficits on social media. *First Draft*, (available at https://firstdraftnews.org:443/long-form-article/under-the-surface-covid-19-vaccine-narratives-misinformation-and-data-deficits-on-social-media/).

35. N. Calleja, A. AbdAllah, N. Abad, N. Ahmed, D. Albarracin, E. Altieri, J. N. Anoko, R. Arcos, A. A. Azlan, J. Bayer, A. Bechmann, S. Bezbaruah, S. C. Briand, I. Brooks, L. M. Bucci, S. Burzo, C. Czerniak, M. D. Domenico, A. G. Dunn, U. K. H. Ecker, L. Espinosa, C. Francois, K. Gradon, A. Gruzd, B. S. Gülgün, R. Haydarov, C. Hurley, S. I. Astuti, A. Ishizumi, N. Johnson, D. J. Restrepo, M. Kajimoto, A. Koyuncu, S. Kulkarni, J. Lamichhane, R. Lewis, A. Mahajan, A. Mandil, E. McAweeney, M. Messer, W. Moy, P. N. Ngamala, T. Nguyen, M. Nunn, S. B. Omer, C. Pagliari, P. Patel, L. Phuong, D. Prybylski, A. Rashidian, E. Rempel, S. Rubinelli, P. Sacco, A. Schneider, K. Shu, M. Smith, H. Sufehmi, V. Tangcharoensathien, R. Terry, N. Thacker, T. Trewinnard, S. Turner, H. Tworek, S. Uakkas, E. Vraga, C. Wardle, H. Wasserman, E. Wilhelm, A. Würz, B. Yau, L. Zhou, T. D. Purnat, A Public Health Research Agenda for Managing Infodemics: Methods and Results of the First WHO Infodemiology Conference. *JMIR Infodemiology*. **1**, e30979 (2021).

36. D. M. J. Lazer, M. A. Baum, Y. Benkler, A. J. Berinsky, K. M. Greenhill, F. Menczer, M. J. Metzger, B. Nyhan, G. Pennycook, D. Rothschild, M. Schudson, S. A. Sloman, C. R. Sunstein, E. A. Thorson, D. J. Watts, J. L. Zittrain, The science of fake news. *Science*. **359**, 1094–1096 (2018).

37. Debunking Handbook 2020 || Databrary, (available at https://nyu.databrary.org/volume/1182).

38. Y. G. Wanless Andrew Gully, Yoel Roth, Abhishek Roy, Joshua A. Tucker, Alicia, Evidence-Based Misinformation Interventions: Challenges and Opportunities for Measurement and Collaboration. *Carnegie Endowment for International Peace*, (available at https://carnegieendowment.org/2023/01/09/evidence-based-misinformation-interventions-challenges-and-opportunities-for-measurement-and-collaboration-pub-88661).

39. glocal, adj. *OED Online*, (available at https://www.oed.com/view/Entry/276090).

40. N. F. Johnson, N. Velásquez, N. J. Restrepo, R. Leahy, N. Gabriel, S. El Oud, M. Zheng, P. Manrique, S. Wuchty, Y. Lupu, The online competition between pro- and anti-vaccination views. *Nature*. **582**, 230–233 (2020).

41. J. Madhusoodanan, Safe space: online groups lift up women in tech. *Nature*. **611**, 839–841 (2022).

42. R. Y. Moon, A. Mathews, R. Oden, R. Carlin, Mothers' Perceptions of the Internet and Social Media as Sources of Parenting and Health Information: Qualitative Study. *Journal of Medical Internet Research*. **21**, e14289 (2019).

43. T. Ammari, S. Schoenebeck, ""Thanks for your interest in our Facebook group, but it's only for dads": Social Roles of Stay-at-Home Dads" in *Proceedings of the 19th ACM Conference on Computer-Supported Cooperative Work & Social Computing* (Association for Computing Machinery, New York, NY, USA, 2016; https://dl.acm.org/doi/10.1145/2818048.2819927), *CSCW '16*, pp. 1363–1375.

44. R. Laws, A. D. Walsh, K. D. Hesketh, K. L. Downing, K. Kuswara, K. J. Campbell, Differences Between Mothers and Fathers of Young Children in Their Use of the Internet to Support Healthy Family Lifestyle Behaviors: Cross-Sectional Study. *J Med Internet Res*. **21**, e11454 (2019).

45. D. R. Forsyth, *Group Dynamics* (Wadsworth Cengage Learning, Belmont, CA, ed. 6th, 2014).

46. M. J. Gelfand, J. R. Harrington, J. C. Jackson, The Strength of Social Norms Across Human Groups. *Perspect Psychol Sci*. **12**, 800–809 (2017).

47. K. H. Tram, S. Saeed, C. Bradley, B. Fox, I. Eshun-Wilson, A. Mody, E. Geng, Deliberation, Dissent, and Distrust: Understanding Distinct Drivers of Coronavirus Disease 2019 Vaccine Hesitancy in the United States. *Clinical Infectious Diseases*. **74**, 1429–1441 (2022).

48. E. Pertwee, C. Simas, H. J. Larson, An epidemic of uncertainty: rumors, conspiracy theories and vaccine hesitancy. *Nat Med*. **28**, 456–459 (2022).

49. I. Freiling, N. M. Krause, D. A. Scheufele, D. Brossard, Believing and sharing misinformation, fact-checks, and accurate information on social media: The role of anxiety during COVID-19. *New Media & Society*. **25**, 141–162 (2023).

50. H. Song, J. So, M. Shim, J. Kim, E. Kim, K. Lee, What message features influence the intention to share misinformation about COVID-19 on social media? The role of efficacy and novelty. *Computers in Human Behavior*. **138**, 107439 (2023).

51. A. Arriagada, F. Ibáñez, "You Need At Least One Picture Daily, if Not, You're Dead": Content Creators and Platform Evolution in the Social Media Ecology. *Social Media + Society*. **6**, 2056305120944624 (2020).

52. Y. A. Kim, M. A. Ahmad, Trust, distrust and lack of confidence of users in online social media-sharing communities. *Knowledge-Based Systems*. **37**, 438–450 (2013).

53. X. Chen, S.-C. J. Sin, Y.-L. Theng, C. S. Lee, Why Students Share Misinformation on Social Media: Motivation, Gender, and Study-level Differences. *The Journal of Academic Librarianship*. **41**, 583–592 (2015).

54. X. Chen, S.-C. J. Sin, Y.-L. Theng, C. S. Lee, "Why Do Social Media Users Share Misinformation?" in *Proceedings of the 15th ACM/IEEE-CS Joint Conference on Digital Libraries* (Association for Computing Machinery, New York, NY, USA, 2015; https://doi.org/10.1145/2756406.2756941), *JCDL '15*, pp. 111–114.

55. D. Centola, J. Becker, D. Brackbill, A. Baronchelli, Experimental evidence for tipping points in social convention. *Science*. **360**, 1116–1119 (2018).

56. A. K. Constantino, Health officials confirm first U.S. case of monkeypox virus this year in Massachusetts. *CNBC* (2022), (available at https://www.cnbc.com/2022/05/19/monkeypox-virus-case-confirmed-in-massachusetts.html).

57. A. Liptak, In 6-to-3 Ruling, Supreme Court Ends Nearly 50 Years of Abortion Rights. *The New York Times* (2022), (available at https://www.nytimes.com/2022/06/24/us/roe-wade-overturned-supreme-court.html).

58. OLCA, The President Signs H.R. 5376, the "Inflation Reduction Act of 2022," (available at https://www.ssa.gov/legislation/legis_bulletin_081622.html).

59. 2022 Midterm Election Calendar - 270toWin. *270toWin.com*, (available at https://www.270towin.com/2022-election-calendar/).

60. M. Jacomy, T. Venturini, S. Heymann, M. Bastian, ForceAtlas2, a Continuous Graph Layout Algorithm for Handy Network Visualization Designed for the Gephi Software. *PLOS ONE*. **9**, e98679 (2014).

61. A. Dance, How one virus can block another, (available at https://www.bbc.com/future/article/20230210-can-you-get-two-viruses-at-the-same-time).

62. S. Dixon, Facebook users by country 2023. *Statista* (2023), (available at https://www.statista.com/statistics/268136/top-15-countries-based-on-number-of-facebook-users/).

63. N. F. Johnson, M. Zheng, Y. Vorobyeva, A. Gabriel, H. Qi, N. Velasquez, P. Manrique, D. Johnson, E. Restrepo, C. Song, S. Wuchty, New online ecology of adversarial aggregates: ISIS and beyond. *Science*. **352**, 1459–1463 (2016).

64. N. F. Johnson, R. Leahy, N. J. Restrepo, N. Velasquez, M. Zheng, P. Manrique, P. Devkota, S. Wuchty, Hidden resilience and adaptive dynamics of the global online hate ecology. *Nature*. **573**, 261–265 (2019).

65. R. F. S. Rhys Leahy, N. J. Restrepo, Y. Lupu, N. F. Johnson, Machine Learning Language Models: Achilles Heel for Social Media Platforms and a Possible Solution. *AAIML*. **01**, 191–202 (2021).

66. C. R. Sunstein, *#Republic: Divided Democracy in the Age of Social Media* (Princeton University Press, NED-New edition., 2018; https://www.jstor.org/stable/j.ctv8xnhtd).

67. World Map of Social Networks. *Vincos - il blog di Vincenzo Cosenza*, (available at https://vincos.it/world-map-of-social-networks/).

68. C. Lampe, J. Vitak, R. Gray, N. Ellison, "Perceptions of facebook's value as an information source" in *Proceedings of the SIGCHI Conference on Human Factors in Computing Systems* (Association for Computing Machinery, New York, NY, USA, 2012; https://doi.org/10.1145/2207676.2208739), *CHI '12*, pp. 3195–3204.

69. A. Silver, L. Matthews, The use of Facebook for information seeking, decision support, and self-organization following a significant disaster. *Information, Communication & Society*. **20**, 1680–1697 (2017).

70. L. Illari, N. F. Johnson, Network resilience in the face of deplatforming: the online anti-vaccination movement and COVID-19 (2022), (available at https://researchshowcase2022-gwu.ipostersessions.com/?s=7F-4B-0F-52-FE-EB-CA-63-04-D1-99-44-89-00-52-E6).